\documentclass{amsart}

\usepackage{latexsym}

\newtheorem{lem}{Lemma}

\newtheorem{prop}{Proposition}

 \def\NN{\hbox{\sf I\kern-.13em\hbox{N}}}
 \def\HH{\hbox{\sf I\kern-.13em\hbox{H}}}
 \def\DD{\hbox{\sf I\kern-.13em\hbox{D}}}
 \def\RR{\hbox{\sf I\kern-.14em\hbox{R}}}
 \def\CC{\hbox{\sf I\kern-.44em\hbox{C}}}
 \def\ZZ{{\hbox{\sf Z\kern-.43emZ}}}
 \def\QQ{\hbox{\sf C\kern -.48emQ}}
 \def\Cc{\hbox{\sf C\kern -.47em {\raise .48ex \hbox{$\scriptscriptstyle |$}}
   \kern-.5em {\raise .48ex \hbox{$\scriptscriptstyle |$}} }}
 \def\Qq{\hbox{\sf Q\kern -.57em {\raise .48ex \hbox{$\scriptscriptstyle |$}}
   \kern-.55em {\raise .48ex \hbox{$\scriptscriptstyle |$}} }}
 \def\BB{\hbox{\sf ]\kern-.5em\hbox{[}}}

\newcommand\e{{\rm e}}
\newcommand{\beq}{\begin{equation}}
\newcommand{\eeq}{\end{equation}}
\newcommand{\R}{{\rm Re}}

\newcommand{\Res}{{\rm Res}}

\title{On the non homogeneous quadratic Bessel zeta function}

\author{M. Spreafico}
\address{ICMC-USP,
Universidade S\~{a}o Paulo, S\~{a}o Carlos, Brasil, mauros@icmc.usp.br}

\begin{document}

\begin{abstract} We study the non homogeneous quadratic
Bessel zeta function $\zeta_{RB}(s,\nu,a)$ defined
as the sum of the square of the positive zeros of
the Bessel function $J_\nu(z)$ plus a
positive constant. In particular,
we give explicit formulas for the main associated zeta invariants,
namely poles and residua, $\zeta_{RB}(0,\nu,a)$ and $\zeta'_{RB}(0,\nu,a)$.
\end{abstract}

\maketitle

\leftline{ Keywords: zeta function, zeta invariant, regularized determinant}

\vskip .2in

From the point of view of differential geometry and mathematical physics,
the Riemann
zeta function appears as the operator zeta function associated to the Laplacian
operator on the line segment \cite{Ros} \cite{RS} \cite{BG} \cite{BO} \cite{Haw}.
A natural generalisation of this setting,
is to consider a Sturm Liouville operator instead, i.e. a singularity at one of the
end points \cite{Cal} \cite{Che1} \cite{Che2} \cite{BS1} \cite{BS2}\cite{Les}.
This leads again to a concrete zeta function,
namely the Bessel zeta function, where the sum is extended on the positive
zeros of the Bessel function $J_\nu(z)$, and reduces for
the opportune choice of $\nu$
to the classical Riemann case. Such a function was first considered and
studied by Stolarsky in \cite{Sto}, where formulas for poles and residua are given,
and more recently by other authors, who calculated the associated zeta invariants by
different methods \cite{AB} \cite{Les}.
In these notes, we study the non homogeneous version of this function. We
determinate his poles and give formulas for the residua. In particular, we
introduce two simple but quite
general methods to calculate
the value of the derivative at the origin, and therefore
the regularized determinant of the associated Sturm-Liouville singular operator
\cite{BFK1} \cite{BFK2} \cite{CQ} \cite{Spr1}.

\vskip .2in

Consider the constant singular Sturm Liouville operator
\[
L_\nu +q^2=-\frac{d^2}{dx^2}+\frac{4\nu^2-1}{4x^2}+q^2,
\]
on the line interval $(0,l]$, with positive real $\nu$ and $q$ (the null cases
can be easily obtained as  limit cases).
$L_\nu$ has the discrete resolution \cite{Gil}
\[
\left\{\lambda_{\nu,n}^2+q^2=\frac{j^2_{\nu, n}}{l^2}+q^2,
\phi_{\nu,n}(x)=\frac{\sqrt{2x}
J_\nu(\lambda_{\nu,n} x )}{lJ_{\nu+1}(j_{\nu,n})}\right\},
\]
where $j_{\nu, n}$ are the positive zeros of the
Bessel function $J_\nu(z)$ ordered in increasing way \cite{Wat}.

For analogy with the Riemann case, we consider the following non homogeneous
quadratic Bessel zeta function defined by:
\[
\zeta_{RB}(s,\nu,a)=\sum_{n=1}^\infty (n_{\nu,n}^2+a^2)^{-\frac{s}{2}},
\]
for $\R (s)>1$, where $\pi n_{\nu,n}=j_{\nu ,n}$ and $a$ is real and positive,
and we study its analytical extension.
We can state our main result for the Bessel zeta function, that comes as a
corollary of the more general result state in Proposition \ref{PP} below
concerning the zeta function associated to the operator $L_\nu+q^2$.

\begin{prop} The function $\zeta_{RB}(s,\nu,a)$ has an analytic extension to the
complex $s$-plane, smooth in $\nu$, up to a discrete set
of simple poles at $s=1,-1,-3, \dots,$
whose residua can be computed using the known asymptotic expansions for the Bessel
functions ( more precisely, they are given by the residua of the function
$z(s,\nu,a,\pi)$ in Proposition \ref{PP} multiplied by 2); in particular
\[
\Res_1\left(\zeta_{RB}(s,\nu,a),s=1)\right)
=1,
\]
\[
\Res_1\left(\zeta_{RB}(s,\nu,a),s=-1)\right)
=-\frac{1}{2\pi^2}\left(\nu^2-\frac{1}{4}-\pi^2 a^2\right).
\]

This extension is regular at $s=0$, and
\[
\Res_0(\zeta_{RB}(s,\nu,a),s=0)
=-\frac{1}{2}\left(\nu+\frac{1}{2}\right),
\]
\[
\Res_0(\zeta'_{RB}(s,\nu,a),s=0)
=-\frac{1}{2}\log\sqrt{2}\pi \frac{I_\nu(\pi a)}{a^\nu}.
\]
\label{p1}
\end{prop}

Notice that the values for the homogeneous case, follow immediately using the series
expansion for the Bessel function $I_\nu(z)$ for small $z$, namely (see \cite{Les})
\[
\zeta_{RB}'(0,\nu,0)=\frac{1}{2}\log
\frac{2^{\nu-\frac{1}{2}}\Gamma(\nu+1)}{\pi^{\nu+1}}.
\]

\hskip .4in

We will study the more general setting, i.e. the function
\[
z(s,\nu,q,l)=\zeta(s;L_\nu+q^2)
=\sum_{n=1}^\infty (\lambda_{\nu,n}^2+q^2)^{-s}.
\]

Convergence of the series for $\R(s)>\frac{1}{2}$ follows from
classical estimates on the zeros of Bessel functions \cite{Wat}, and
$\zeta_{RB}(2s,\nu,a)=z(s,\nu,a,\pi)$.

\vskip .2in

We present here two approaches to calculate the zeta invariants associated to
$z(s,\nu,q,l)$. In the first, we produce an analytic representation that can be
effectively used to get all the invariants; this will be particularly useful to
generalise the method for the calculation of the zeta invariants associated to a
disk or to a cone \cite{Spr2}.  In the second approach, we give a very general
lemma (Lemma \ref{lll}) to deal with regularized products,
and we apply it to calculate $z(0,\nu,q,l)$ and $z'(0,\nu,q,l)$.

\vskip .2in

Let start with the first approach.
Using Mellin transform \cite{Gil} \cite{LM}, we get the analytic
representation
\[
z(s,\nu,q,l)=\frac{1}{\Gamma(s)}\int_0^\infty t^{s-1} f(t,\nu,q,l) dt,
\]
where the trace of the heat operator is
\[
f(t,\nu,q,l)={\rm Tr}\e^{-tL_\nu}
=\sum_{n=1}^\infty \e^{-(\lambda_{\nu,n}^2+q^2)},
\]
and from this the complex representation
\[
f(t,\nu,q,l)=\frac{1}{2\pi i}\int_{\Lambda_c} \e^{-\lambda t}
R(\lambda,\nu,q,l) d\lambda,
\]
where the contour is
$\Lambda_c=\{\lambda\in\CC~|~|\arg(\lambda-c)|=\pi/4\}$,
for some $0<c<q^2$, and the trace of the resolvent is
\[
R(\lambda,\nu,q,l)=\sum_{n=1}^\infty
\frac{1}{\lambda-(\lambda_{\nu,n}^2+q^2)}.
\]

We now observe that it is easy to express such function in terms of special
functions. In fact, taking logarithmic derivative of the infinite product
representation of the Bessel function $I_\nu(z)$, we get \cite{Wat}:

\begin{lem}
\[
R(\lambda,\nu,q,l)=\frac{\nu}{2z^2}-\frac{1}{2z}\frac{d}{dz}\log I_\nu(l z).
\]
\label{l1}
\end{lem}

Here, $z=\sqrt{q^2-\lambda}$, we set
$\arg(q^2-\lambda)=0$ on the line $(-\infty, q^2)$ and we fix
the sector $s_+=\{z\in\CC~|~|\arg z|<\pi/2\}$ for $z$.

At this point it is worth observing that information about poles and residua of
$z(s,\nu,q,l)$ can be obtained using the representation introduced, asymptotics
expansions for Bessel functions and classical arguments \cite{Gil} \cite{LM}.
This is an easy way for producing the results
relative to the so called 'constant case' when studying regular singular operators
\cite{Cal} \cite{BS1} \cite{BS2}. More precisely, and for completeness,
we can state the following
results:

\begin{lem} For small $t$,
\[
f(t,\nu,q,l)=\sum_{i=0}^I a_i(\nu,q,l) t^{(i-1)/2} +O(t^{I/2})=
\]
\[
=\frac{l}{2\sqrt{\pi}}
t^{-1/2}-\frac{1}{2}\left(\nu+\frac{1}{2}\right)
+\frac{1}{2l\sqrt{\pi}}
\left(\nu^2-\frac{1}{4}-l^2q^2\right)t^{1/2}
+O(t),
\]
where
\[
a_{2i}(\nu,q,l)=\frac{(-1)^i}{i!}\frac{l q^{2i}}{2\sqrt{\pi}}
-\sum_{j=0}^\infty\sum_{k=1,k+2j=2i-1}^\infty
\frac{(-1)^{j+k}}{2^k j! k!}\frac{q^{2j}}{l^k \Gamma\left(\frac{k}{2}\right)}
\frac{\Gamma\left(\nu+k+\frac{1}{2}\right)}{\Gamma\left(\nu-k+\frac{1}{2}\right)},
\]
\[
a_{2i+1}(\nu,q,l)=\frac{(-1)^{i+1}\left(\nu-\frac{1}{2}\right)q^{2i}}{2i!}
-\sum_{j=0}^\infty\sum_{k=1,k+2j=2i}^\infty
\frac{(-1)^{j+k}}{2^k j! k!}\frac{q^{2j}}{l^k \Gamma\left(\frac{k}{2}\right)}
\frac{\Gamma\left(\nu+k+\frac{1}{2}\right)}{\Gamma\left(\nu-k+\frac{1}{2}\right)}.
\]
\end{lem}

\begin{prop} The function $z(s,\nu,q,l)$ has an analytic extension to the
whole complex $s$-plane, smooth in $\nu$ and $q$, up to a discrete set
of simple poles at $s=\frac{1}{2}, -\frac{1}{2}, -\frac{3}{2}, \dots,$
with residua
\[
\Res_1\left(z(s,\nu,q,l),s=\frac{1}{2}-k\right)
=\frac{a_{2k}(\nu,q,l)}{\Gamma\left(\frac{1}{2}-k\right)},~~~k=0,1,2,\dots.
\]

This extension is regular at $s=0$, and
\[
\Res_0(z(s,\nu,q,l),s=0)
=-\frac{1}{2}\left(\nu+\frac{1}{2}\right).
\]

\label{p2}
\end{prop}

Notice that the contribution of the non homogeneity term $q^2$ is equally shared
among all the poles; in other words, the homogeneous Bessel zeta function has the
same poles, but with (possibly) different residua (see \cite{Sto}).

The information available is not enough to deal with the
derivative, that is an harder point; though, we introduce the following quite
more general purpose result.

\begin{lem}
Suppose the zeta function $z(s,x)=\sum_{n=1}^\infty a_n(x)^{-s}$
has the following representation (everything smooth in $x$):
\[
z(s,x)
=\frac{1}{\Gamma(s)}\int_0^\infty t^{s-1} \frac{1}{2\pi i}
\int_{\Lambda_c} \e^{-\lambda t} R(\lambda,x)d\lambda dt,
\]
where the contour is as above, and the there is a primitive function $-T(\lambda,x)$
for the function $R(\lambda,x)=-\frac{d}{d\lambda}T(\lambda,x)$, satisfying the
following properties:

\begin{itemize}

\item[]{(a)} $T$ is analytic near $\lambda=0$,

\item[]{(b)} for large $\lambda$ and fixed $x$,
there is an asymptotic expansion
\[
T(\lambda,x)=\dots+A(x)\log(-\lambda)+B(x)+\dots.
\]

\end{itemize}

Then, $z(s,x)$ can be analytically extended at $s=0$ and
\[
z(0,x)=-A(x),
\]
\[
z'(0,x)=-B(x)+T(0,x).
\]
\label{p3}
\end{lem}

\noindent\underline{Proof} Integrating by part, first in $\lambda$ and hence in $t$,
the given complex representation for $z(s,x)$, we get
\[
z(s,x)
=\frac{s}{\Gamma(s)}\int_0^\infty t^{s-1} \frac{1}{2\pi i}
\int_{\Lambda_c} \frac{\e^{-\lambda t}}{-\lambda} T(\lambda,x)d\lambda dt.
\]

At this point, it is convenient to isolate the constant part of $T$ writing
\[
T(\lambda,x)=T_1(\lambda,x)+T_0(x);
\]
in fact, it is clear that $T_0(x)$ gives no contribution, since
\[
\int_{\Lambda_c} \frac{\e^{-\lambda t}}{-\lambda} T_0(x)d\lambda =0.
\]

Next, since by definition the
zeta function is well defined for large $s$,
\[
z(s,x)
=\frac{s^2}{\Gamma(s+1)}\int_0^1 t^{s-1} \frac{1}{2\pi i}
\int_{\Lambda_c} \frac{\e^{-\lambda t}}{-\lambda} T_1(\lambda,x)d\lambda dt
+s^2 f(s),
\]
where $f$ is regular near $s=0$. Because of the pole at $\lambda=0$, we
have to split the complex integral as follows to use the expansion for
large $\lambda$ (small $t$):
\[
\int_{\Lambda_c}=\int_{\Lambda_{-c}}-\int_{C_c}
=\int_{\Lambda_{-c}}+T_1(0,x),
\]
where $C_c$ is a circle around the origin of radius $c$.
Moreover, by assumption (b),
\[
\int_{\Lambda_{-c}} \frac{\e^{-\lambda t}}{-\lambda}
T_1(\lambda,x)d\lambda dt
=\int_{\Lambda_{-c}} \frac{\e^{-\lambda }}{-\lambda}
T_1(\lambda/t,\nu,q)d\lambda dt=
\]
\[
=\dots +\gamma A(x)+A(x)\ln(-\lambda)+B(x)+\dots,
\]
where we have explicit-ed only the relevant part. This means that we can
write
\[
z(s,x)=\frac{s}{\Gamma(s+1)}\left[ \gamma
A(x)-B(x)-\frac{1}{s}A(x)+T_1(0,x)\right]
+s^2 g(s),
\]
where again $g$ is regular near $s=0$, and, from that, the thesis
follows at once.
$\Box$

\hskip .4in

Applying this argument to the function $z(s,\nu,q,l)$, we have
\[
T(\lambda,\nu,q,l)=\nu\log lz-\log I_\nu(lz)-\log 2^\nu \Gamma(\nu+1),
\]
and hence we compute
\[
A(\nu,q,l)=\frac{1}{2}\left(\nu+\frac{1}{2}\right),
\hskip .1in B(\nu,q,l)=\frac{1}{2}\log 2\pi+\left(\nu+\frac{1}{2}\right)\log l
-\log2^\nu\Gamma(\nu+1),
\]
\[
T(0,\nu,q,l)=\nu\log lq -\log I_\nu(lq)-\log 2^\nu \Gamma(\nu+1).
\]

This gives:
\[
{\rm Res}_0(z(s,\nu,q,l),s=0)=-\frac{1}{2}\left(\nu+\frac{1}{2}\right),
\]
and proves the following:

\begin{prop}
\[
{\rm Res}_0(z'(s,\nu,q,l),s=0)=-\log\sqrt{2\pi l}\frac{I_\nu(lq)}{q^\nu}.
\]
\label{PP}
\end{prop}

Let's turn to the second approach. First, we have the following:

\begin{lem} Suppose two sequences $a_n$ and $b_n$, $n=1,2,3,\dots$, of
real positive numbers are given, and satisfy the following conditions:

\begin{itemize}

\item[(A)] there are real $s_a$, $s_b$ such that the two series
\[
\zeta_a(s)=\sum_{n=1}^\infty a_n^{-s},~~~~~~~~
\zeta_b(s)=\sum_{n=1}^\infty b_n^{-s},
\]
converge for $\R(s)>s_a, s_b$ respectively;

\item[(B)] the zeta function $\zeta_b(s)$ has an analytic extension at
$s=0$

\item[(C)] $|a_n-b_n|<K$ definitely for some constant $K$,

\item[(D)] $s_b<1$.

\end{itemize}

Then, the zeta function $\zeta_a(s)$ has an analytic extension at $s=0$ with
$\zeta_a(0)=\zeta_b(0)$, the infinite product
\[
\prod_{n=1}^\infty \frac{a_n}{b_n}=C
\]
converges absolutely, and
\[
\zeta_a'(0)=\zeta_b'(0)+\log C.
\]
\label{lll}
\end{lem}

\noindent\underline{Proof}
Let $c_n=\frac{a_n}{b_n}-1$. Then, $|c_n|<Kb_n^{-1}$ by (B) and (C) and hence
$\sum_{n=1}^\infty c_n$ converges absolutely and so does the infinite product.
We can assume $c_n<1$, and hence
\[
\zeta_a(s)
=\zeta_b(s)+\sum_{k=1}^\infty \left(\begin{array}{c}-s\\k\end{array}\right)
\sum_{n=1}^\infty b_n^{-s} c_n^k.
\]

Now, the series
\[
\sum_{n=1}^\infty b_n^{-s} c_n^k,
\]
converges in a neighbourhood of $s=0$ for each $k$, by conditions (C) and
(D). This means that $\zeta_a$ can be analitycally extended to $s=0$ by
using the extension of $\zeta_b$. In particular, evaluating at $s=0$ we
get $\zeta_a(0)=\zeta_b(0)$ and
\[
\zeta_a'(0)=\zeta_b'(0)
+\left. \sum_{k=1}^\infty
\frac{d}{ds}\left(\begin{array}{c}-s\\k\end{array}\right)
\sum_{n=1}^\infty b_n^{-s} c_n^k\right|_{s=0}
+\left. \sum_{k=1}^\infty
\left(\begin{array}{c}-s\\k\end{array}\right)
\sum_{n=1}^\infty \log b_n b_n^{-s} c_n^k\right|_{s=0}=
\]
\[
=\zeta_b'(0)
+ \sum_{k=1}^\infty \frac{(-1)^{k-1}}{k} \sum_{n=1}^\infty  c_n^k
=\zeta_b'(0)+\sum_{n=1}^\infty  \log(1+c_n)
=\zeta_b'(0)+ \log C.
\]
$\Box$

We can apply this lemma to the present case as follows. Let
$a_n=\lambda_{\nu,n}^2+q^2$,
$b_n=\frac{\pi^2}{l^2}\left[n+\frac{1}{2}\left(\nu-\frac{1}{2}\right)\right]^2$;
then, all the assumptions of Lemma \ref{lll} are satisfied, and $\zeta_b(s)$ is the
Hurwitz zeta function
$\zeta_H \left(2s,\frac{1}{2}\left(\nu+\frac{3}{2}\right)\right)$. Thus,
\[
\zeta_b'(0)= 2\log\Gamma\left(\frac{1}{2}\left(\nu+\frac{3}{2}\right)\right)
-\log 2\pi,
\]
and we can compute (where $u=\frac{1}{2}\left(\nu-\frac{1}{2}\right)$):
\[
\frac{1}{C} =\lim_{z\to 0^+}
\frac{\prod_{n=1}^\infty\left(1+\frac{l^2}{z^2(j_{\nu,n}^2+q^2l^2)}\right)}
{\prod_{n=1}^\infty\left(1+\frac{l^2}{\pi^2 z^2(n+u)^2)}\right)}
=\lim_{z\to 0^+}
\frac{\prod_{n=1}^\infty\left(1+\frac{l^2(1+q^2z^2)}{z^2j_{\nu,n}^2}\right)}
{\prod_{n=1}^\infty\left(1+\frac{l^2q^2}{j_{\nu,n}^2}\right)
\left(1+\frac{l^2}{\pi^2 z^2(n+u)^2)}\right)}=
\]
\[
=\lim_{z\to 0^+} \frac{z^\nu q^\nu
I_\nu\left(\frac{l\sqrt{1+q^2z^2}}{z}\right)
\left|\Gamma\left(u+1+i\frac{l}{\pi z}\right)\right|^2} {I_\nu(lq)
\Gamma^2(u+1)} =\frac{\sqrt{2} (lq)^\nu}{\pi^\nu I_\nu(lq)
\Gamma^2\left(\frac{1}{2}\left(\nu+\frac{3}{2}\right)\right)}.
\]


\begin{thebibliography}{99}

\bibitem{AB} A. Actor and I. Bender, {\em The zeta function constructed from the
zeros of the Bessel function}, J. Phys. A 29 (1996) 6555-6580;

\bibitem{ABP} M. Atiyah, R. Bott and V.K. Patodi, {\em On the heat
equation and the index theorem}, Invenct. Math. 19 (1973) 279-330;

\bibitem{BFK1} D. Burghelea, L. Friedlander and T. Kappeler, {\em On
the determinant of elliptic differential and finite difference
operators in  vector bundles over $S^1$}, Comm. Math. Phys. 138 (1991) 1-18;


\bibitem{BFK2} D. Burghelea, L. Friedlander and T. Kappeler, {\em On
the determinant of elliptic boundary value problems on a line segment},;

\bibitem{BG} T.P Bransom and P.B. Gilkey, {\em The functional determinant of a
four-dimensional boundary value problem}, Trans. Amer. Math. Soc. 344 (1994)
479-531;

\bibitem{BO} T.P. Bransom and B. Orsted, {\em Conformal geometry and local
invariants}, Diff. Geom. Appl. 1 (1991) 279-308;

\bibitem{BS1} J. Bruning and R. Seeley, {\em Regular singular
asymptotics}, Adv. in Math. 58 (1985) 133-148;

\bibitem{BS2} J. Bruning and R. Seeley, {\em The resolvent expansion for
second order regular singular operators}, J. of Funct. An. 73 (1988)
369-415;

\bibitem{Cal} C. Callias, {\em The heat equation with singular
coefficients}, Comm. Math. Phys. 88 (1983) 357-385;

\bibitem{Che1} J. Cheeger, {\em On the spectral geometry of spaces with
conical singularities}, Proc. Nat. Acad. Sci. 76 (1979) 2103-2106;

\bibitem{Che2} J. Cheeger, {\em Spectral geometry of singular
riemannian spaces}, J. Diff. Geom. 18, (1983) 575-657;

\bibitem{CQ} J. Choi and J.R. Quine, {\em Zeta regularized products and
functional determinants on spheres}, Rocky Mount. Jour. Math. 26 (1996)
719-729;


\bibitem{Gil} P.B. Gilkey, {\em Invariance theory, the heat equation, and
the Atiyah-Singer index theorem}, second edition 1995 CRC press;


\bibitem{Haw} S.W. Hawking, {\em Zeta function regularization of path
integrals in curved space time}, CMP 55 (1977) 133-148;


\bibitem{LM} H.B. Lawson and M.L. Michelsohn, {\em Spin geometry}, Princeton
Math. Series 38 (1989);

\bibitem{Les} M. Lesh, {\em Determinants of regular singular
Sturm-Liouville operators}, Math. Nachr. 194 (1998) 139-170;


\bibitem{RS} D.B. Ray and I.M. Singer, {\em R-torsion and the Laplacian
on riemannian manifolds}, Adv. Math. 7 (1974) 145-210;

\bibitem{Ros} S. Rosenberg, {\em The Laplacian on a riemannian
manifold}, LMSST 31;


\bibitem{Spr1} M. Spreafico, {\em Zeta function and regularized determinant on
projective spaces}, in publication on the Rocky Mount. J. Math. (2001);

\bibitem{Spr2} M. Spreafico, {\em Zeta invariants on a disk and on a cone},
submitted to PAMS (2002);

\bibitem{Sto} K.B. Stolarsky, {\em Singularities of Bessel-zeta functions and
Hawkins' polynomials}, Mathematika 32 (1985) 96-103;


\bibitem{Wat} G.N. Watson, {\em A treatise on the theory of Bessel
functions}, Cambridge University Press 1922.


\end{thebibliography}
\end{document}